\documentclass{Interspeech2024}

\interspeechcameraready

\usepackage{bm}
\usepackage{multirow}
\usepackage{algorithm} 
\usepackage{algpseudocode} 
\usepackage{url}

\usepackage{breakurl}
\usepackage[dvipsnames]{xcolor}
\usepackage{xcolor}
\usepackage{amsfonts}
\usepackage{booktabs} 
\usepackage{pifont}
\usepackage{ragged2e}
\usepackage{hyperref}
\usepackage{subcaption}
\usepackage{siunitx}
\ninept 
\title{ Learning Fine-Grained Controllability on Speech Generation via Efficient Fine-Tuning }

\name[affiliation={,1}]{Chung-Ming}{Chien$^\dagger$}
\name[affiliation={2}]{Andros}{Tjandra}
\name[affiliation={2}]{Apoorv}{Vyas}
\name[affiliation={2}]{Matt}{Le}
\name[affiliation={2}]{Bowen}{Shi}
\name[affiliation={2}]{Wei-Ning}{Hsu}

\address{
  $^1$Toyota Technological Institute at Chicago, USA \qquad
  $^2$AI at Meta, USA}

\keywords{Speech generation, fine-grained conditioning, efficient fine-tuning}

\PassOptionsToPackage{hyphens}{url}

\def\check{\ding{51}}
\def\cross{\ding{55}}

\newcommand\blfootnote[1]{%
  \begingroup
  \renewcommand\thefootnote{}\footnote{#1}%
  \addtocounter{footnote}{-1}%
  \endgroup
}

\newif\ifdraft
\draftfalse

\definecolor{dkbrown}{RGB}{170,50,0}

\begin{document}

\maketitle

\blfootnote{$^\dagger$Work done during an internship at Meta.}
\vspace{-3mm}
\begin{abstract}
\vspace{-1mm}
As the scale of generative models continues to grow, efficient reuse and adaptation of pre-trained models have become crucial considerations. 
In this work, we propose Voicebox Adapter, a novel approach that integrates fine-grained conditions into a pre-trained Voicebox speech generation model using a cross-attention module.
To ensure a smooth integration of newly added modules with pre-trained ones, we explore various efficient fine-tuning approaches.
Our experiment shows that the LoRA with bias-tuning configuration yields the best performance, enhancing controllability without compromising speech quality.
Across three fine-grained conditional generation tasks, we demonstrate the effectiveness and resource efficiency of Voicebox Adapter. 
Follow-up experiments further highlight the robustness of Voicebox Adapter across diverse data setups.
\end{abstract}

\vspace{-2mm}
\section{Introduction}
\vspace{-1mm}
\label{sec:intro}
Large-scale speech pre-training has demonstrated remarkable competence across various applications~\cite{yang2021superb, mohamed2022self}.
In contrast to discriminative pre-training~\cite{baevski2020wav2vec, hsu2021hubert}, which aims to acquire speech representations beneficial for downstream tasks, generative pre-training~\cite{liu2024generative} directly learns the data distribution from speech corpora~\cite{lakhotia2021on, borsos2023audio}.
Through pre-training on diverse data, speech generation models learn the distribution of speech of different styles, speakers, and various transient events, including pauses, stress, and non-verbal vocalizations such as laughter~\cite{nguyen2023generative}.

Recent speech generation models have demonstrated one-shot capabilities, allowing control over the speaker and style of generated speech with a short speech prompt~\cite{wang2023neural, le2023voicebox}.
We term this form of controlled generation as \textit{global control}, where the speaker and style remain consistent throughout the entire generated utterance. 
In contrast, \textit{fine-grained control} involves adding conditioning only to specific parts of the generated utterance, such as emphasizing certain words or introducing pauses at specific points.
Despite the intrinsic global controllability observed in many generative pre-training methods~\cite{wang2023neural, le2023voicebox}, the exploration of post-hoc integration of fine-grained controllability into pre-trained speech generation models remains limited.

In this work, we propose Voicebox Adapter, a method to integrate fine-grained controllability into Voicebox~\cite{le2023voicebox}, a text-conditioned speech generation model. 
We explore three fine-grained conditions: punctuation, emphasis, and laughter. 
We hypothesize that these fine-grained vocalizations are inherently learned during pre-training, but the absence of fine-grained conditioning mechanisms in Voicebox restricts its ability to generate speech with precise fine-grained vocalizations.
To address this limitation, we introduce cross-attention modules to the Transformer layers of the pre-trained Voicebox to extract and integrate fine-grained condition information.
Additionally, we employ parameter-efficient adaptation methods to seamlessly connect the pre-trained parameters with the new modules.
Experimental results demonstrate that Voicebox Adapter achieves performance comparable to fine-tuning the entire model, with adapter parameters comprising a small percentage of the model.

Our contributions are as follows:
(1) we propose Voicebox Adapter, which augments Voicebox, a pre-trained speech generation model, with fine-grained controllability;
(2) we explore different efficient fine-tuning methods to bridge the gap between pre-trained parameters and new fine-grained conditioning modules;
(3) we show that Voicebox Adapter can generalize across various fine-grained conditions, attaining performance comparable to that achieved by fine-tuning the entire model with significantly fewer fine-tuned parameters;
(4) we conduct experiments using varying amounts of fine-tuning data and different hidden dimension sizes, analyzing the performance of Voicebox Adapter under different setups.





\vspace{-2mm}
\section{Related works}
\vspace{-1mm}
\subsection{Adaptive fine-grained controllable speech generation}
\vspace{-1mm}
Adaptation has become an important research topic for speech generation. 
A widely adopted approach involves the initial pre-training of a speech generation model on a large and diverse dataset, followed by fine-tuning a specific subset of parameters with a smaller target dataset~\cite{chen2019sample}.
Adaptive methods have shown great success in global speaker~\cite{chen2019sample, chen2021adaspeech} and style control~\cite{yan2021adaspeech}, and can also provide fine-grained controllability~\cite{huang2022generspeech}.
In this work, our goal is to develop a comprehensive adaptive framework that transcends task-specific design considerations~\cite{yan2021adaspeech} and can be applied to various fine-grained conditions.
\vspace{-1mm}
\subsection{Efficient fine-tuning for Transformers}
\vspace{-1mm}
The growing size of language models has made efficient fine-tuning of Transformer models an increasingly important research topic~\cite{houlsby2019parameter}.
Extensive studies have been made to explore the use of adapters~\cite{houlsby2019parameter} and Low-Rank Adaptation (LoRA)~\cite{hu2022lora} across diverse tasks.
Instead of limiting fine-tuning only to newly added modules, recent research also advocates for unlocking specific pre-trained parameters, such as the normalization, bias, and scale of linear layers~\cite{gao2023llama}.
In this paper, we systematically investigate various efficient fine-tuning strategies applied to the Transformer layers of the Voicebox model.

\vspace{-2mm}
\section{Background}
\vspace{-1mm}
Voicebox~\cite{le2023voicebox} is a speech generation framework consisting of a duration model and an acoustic model, both with a Transformer architecture.
Given a phoneme sequence as input, the duration model is trained to predict the duration of each phoneme using an $L_1$ regression loss.
The flow-matching-based~\cite{lipman2023flow} acoustic model defines a vector field to transform a Gaussian prior $p(x)$ into the real distribution of Mel spectrograms $q(x)$.

Let $z_p$ be a time-aligned phoneme embedding sequence,\footnote{During training, a forced-alignment tool is used to obtain the alignment between the phonemes and the ground-truth Mel spectrogram. During inference, the duration predicted by the duration model is used.} and $x_1$ be the associated Mel spectrogram sampled from the real data distribution $q(x)$.
The acoustic model defines a time-dependent vector field $v_t$, which is used to construct a flow $\phi_t$ as described by the differential equation:
\vspace{-0.5mm}
\footnotesize
\begin{equation}
    \frac{d \phi_t(x)}{dt} = v_t(\phi_t(x), m(x_1), z_p; \theta)
    \label{eq:ODE}
\end{equation}
\normalsize
Here, $\theta$ represents the model parameters, $t \in [0, 1]$ is the time parameter, and $m(\cdot)$ is a mask function applied to the Mel spectrogram $x_1$. 
Given $x_0$ sampled from the Gaussian prior $p(x)$, an ODE solver can be used to evaluate $\phi_1(x_0)$ with the initial condition $\phi_0(x_0) = x_0$.

The training objective is to align the flow-transformed distribution $p_1 (x) = p(\phi_1^{-1}(x)) \det [\frac{\partial{\phi_1^{-1}(x)}}{\partial x} (x)]$ with the real data distribution $q(x)$. 
To achieve this, the acoustic model is trained to minimize the loss:
\vspace{-0.5mm}
\footnotesize
\begin{multline}
    \mathbb{E}_{t, (x_1, z_p) \sim q, x_0 \sim p} ||v_t(\psi_t(x_0, x_1), m(x_1), z_p;\theta) - \\
    (x_1 - (1-\sigma_{min})x_0)||^2
\end{multline}
\normalsize
with $\psi_t(x_0, x_1) = (1 - (1 - \sigma_{min}) t) x_0 + t x_1, \sigma_{min}=10^{-5}$.
During training, $t$ is uniformly sampled from $[0, 1]$ and is encoded as a sinusoidal positional embedding, which is concatenated with the phoneme embedding $z_p$, the masked spectrogram $m(x_1)$,\footnote{During training, the mask function $m(\cdot)$ randomly masks out all or parts of the frames in the ground-truth Mel spectrogram $x_1$.}  and the sampled $\psi_t(x_0, x_1)$ as the model input. 

Sampling from the learned audio distribution $p_1(x|m(x_1), z_p)$ starts with a noise $x_0$ sampled  from the Gaussian prior $p(x)$, followed by solving the ODE in equation~(\ref{eq:ODE}) to obtain $\phi_1(x_0)$.
During inference, we can choose to mask out all of $x_1$ for zero-shot text-to-speech or mask parts of $x_1$ to provide additional information to the Voicebox model.

\vspace{-2mm}
\section{Proposed method}
\vspace{-1mm}
\subsection{Voicebox Adapter}
\vspace{-1mm}
Voicebox Adapter extends a pre-trained Voicebox model by incorporating additional modules to handle fine-grained conditions. 
Let $z_f$ be the fine-grained condition, and $\theta'$ denote the parameters of the new fine-grained conditioning modules. 
The vector field modeled by the acoustic model is redefined as:
\vspace{-1mm}
\footnotesize
\begin{equation}
    \frac{d \phi_t(x)}{dt} = v_t(\phi_t(x), m(x_1), z_p, z_f; \theta, \theta')
\end{equation}
\normalsize
During fine-tuning, we keep the pre-trained parameters $\theta$ frozen and solely optimize the new parameters $\theta'$ with the loss:\footnote{Bias-tuning (see section \ref{ssec:method_efficient} for details) is an exception, where the pre-trained LayerNorm layers remain trainable during fine-tuning.}
\vspace{-1mm}
\footnotesize
\begin{multline}
    \mathbb{E}_{t, (x_1, z_p, z_f) \sim q, x_0 \sim p} 
    ||v_t(\psi_t(x_0, x_1), m(x_1), z_p, z_f; \theta, \theta') - \\
    (x_1 - (1-\sigma_{min})x_0)||^2
\end{multline}
\normalsize

\begin{figure}[t]
    \centering 
    \includegraphics[width=0.9\linewidth]{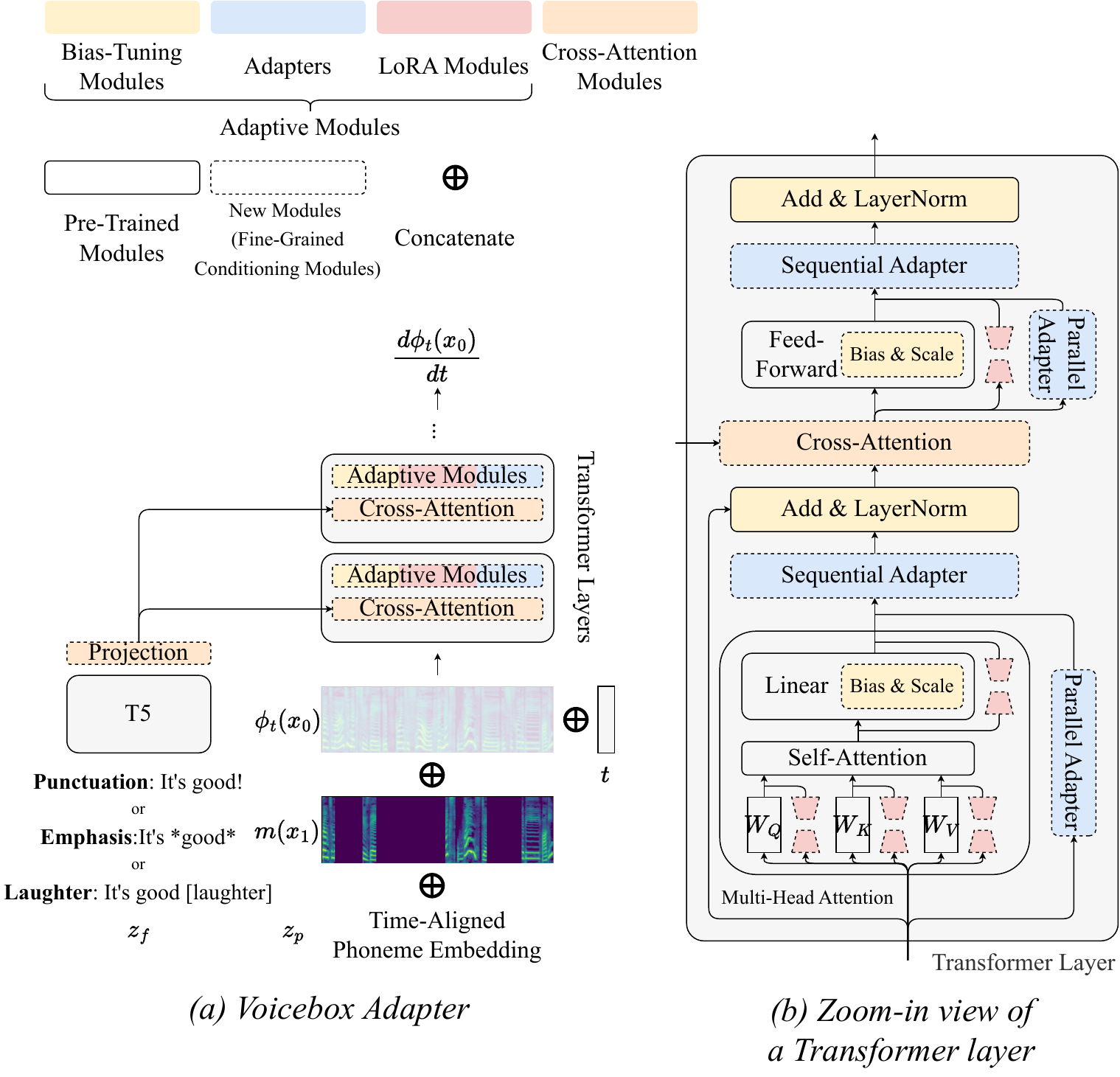}
    \vspace{-3mm}
    \caption{The model architecture of Voicebox Adapter, and a zoom-in view of a Transformer layer.}
    \label{fig:model_arch}
    \vspace{-6mm}
\end{figure}

An illustration of the acoustic model architecture is provided in Fig.\ref{fig:model_arch}(a). 
The newly introduced modules comprise a frozen T5 encoder~\cite{raffel2020exploring}, a trainable linear projection layer, as well as cross-attention modules and adaptive modules within the Transformer stack. 
The T5 model takes fine-grained conditions as inputs and outputs a vector sequence, which is subsequently projected into a 768-dimensional space. 
Cross-attention modules attend to the projected vector sequence to integrate fine-grained conditions into the pre-trained model. 
Within each Transformer layer, parameter-efficient adaptive modules are used to ensure smooth integration of extracted information with the hidden features in the pre-trained Voicebox acoustic model.
As Fig.~\ref{fig:model_arch}(a) shows, the fine-grained conditions are formulated as transcripts with special annotations. 
This framework can potentially be applied to various fine-grained conditioning tasks, provided that the condition can be processed by the T5 encoder.


The implementation of the duration model resembles that of the acoustic model, where the frozen T5 encoder, projection layer, cross-attention modules, and adaptive modules are incorporated to provide fine-grained conditions for duration prediction.
During fine-tuning, we freeze the pre-trained modules and only optimize the new modules as done in the acoustic model.
\vspace{-2mm}
\subsection{Efficient fine-tuning methods}
\vspace{-1.5mm}
\label{ssec:method_efficient}
To facilitate the integration of fine-grained conditions, we explore several efficient fine-tuning strategies, shown in Fig.~\ref{fig:model_arch}(b).
\begin{itemize}
    \item Adapters: We follow the configurations of parallel and sequential adapters in prior work~\cite{he2022towards} and add them to the self-attention and feed-forward layers of each Transformer layer.
    \item LoRA: We explore two different configurations. 
    By default, we apply LoRA to the input projection matrices of the self-attention module.
    We also experiment with adding LoRA to every linear layer in the model, as prior work suggests~\cite{dettmers2023q}.
    \item Bias-tuning: 
    Following the approach of Llama Adapter V2~\cite{gao2023llama}, we apply bias-tuning as an optional add-on to LoRA fine-tuning. 
    In our implementation, bias vector $\mathbf{b}$ and scale vector $\mathbf{s}$ are used to modify the output of every linear layer ($\odot$ means element-wise product):
    \vspace{-1.5mm}
    \begin{equation}
        Linear_{bias-tuning}(\mathbf{x}) = (Linear(\mathbf{x}) +\mathbf{b}) \odot \mathbf{s}
    \vspace{-1.5mm}
    \end{equation}
    During fine-tuning, we optimize the bias and scale vectors jointly with the LayerNorm parameters while keeping the linear layer and other components frozen.
\end{itemize}
For all the adaptive modules, we employ zero-initialization~\cite{liu2024generative} to ensure a smooth start to the fine-tuning process. 


\vspace{-2mm}
\section{Experimental setup}
\vspace{-1mm}
\subsection{Pre-training details}
\vspace{-1.5mm}
Our pre-trained Voicebox acoustic model comprises 12 Transformer layers with a model dimension of 768, and the hidden size of the feed-forward layers is 3,072. 
The model has 12 attention heads. 
We train the model for 750k updates on 32 GPUs, with a batch size of 8k tokens per GPU. 
The duration model has a similar architecture but has only 8 Transformer layers, a model dimension of 512, and a feed-forward hidden size of 2,048, and is trained for 400k updates on 32 GPUs. 
Other details of the models follow from the original setup of Voicebox~\cite{le2023voicebox}.

For pre-training, we utilize a combination of three English datasets: a 60k-hour audiobook reading-style dataset, a 100-hour podcast recording conversation-style dataset, and a 1.7k-hour telephone conversation-style dataset. 
These datasets collectively have over 20k speakers, providing wide coverage of speaking styles. 
To address the unbalanced size of the datasets, we implement data re-sampling to ensure equal exposure to the three datasets. 
We consider audio $x$ to be an 80-dimensional log-scaled Mel spectrogram extracted with a 40\unit{ms} window at a 100\unit{Hz} frame rate, which can be converted to raw waveform with a HiFi-GAN vocoder~\cite{kong2020hifi}.
\vspace{-2.5mm}
\subsection{Fine-grained conditional fine-tuning setup}
\vspace{-1mm}
The cross-attention module for integrating fine-grained conditions has 12 attention heads, each having a hidden dimension of 64.
The configurations of the adaptive modules are as follows:
\begin{itemize}
    \item Adapters: Sequential and parallel adapters are both 2-layer ReLU-activated feed-forward layers with a hidden size of 64.
    \item LoRA: We set the rank (also known as the hidden dimension) $r$ and the scaling parameter $\alpha$ of LoRA both to 64 and use a dropout rate of 0.05.
    \item Bias-tuning: The weight and bias in the pre-trained LayerNorm layers and the newly introduced scale and bias vectors in linear layers are trained.
\end{itemize}
During fine-tuning, only the linear layer on top of the T5 encoder, the cross-attention, and the adaptive modules are trained.
Other optimization setups remain identical to our pre-training setup, except for fewer training updates (see below).

We use different datasets for each fine-grained conditional fine-tuning task. 
For the punctuation task, we use a 550-hour audiobook dataset with punctuated cased transcription.\footnote{Practically, we select the 8 most common punctuation marks (commas, periods, question marks, exclamation marks, colons, semicolons, hyphens, and quotes) for this task and remove other punctuation marks.}\footnotetext[5]{For the punctuation task, we first compute the $F_1$ scores for each type of punctuation mark, and then report the micro-average across the 8 punctuation marks considered.}
For the emphasis task, we use a 20-hour expressive speech dataset with word emphasis annotations in the transcription.
For the laughter task, we use a 250-hour conversation dataset with special annotations for laughter.
The acoustic model is fine-tuned for 50k updates and the duration model is fine-tuned for 100k updates in the punctuation and laughter tasks.
In the emphasis task, due to limited fine-tuning data, we reduce the fine-tuning updates for the acoustic and duration models to 30k/50k, respectively.
\vspace{-2.5mm}
\subsection{Inference setup}
\vspace{-1mm}
The primary focus of Voicebox Adapter is on the zero-shot text-to-speech setup, where we mask out all content in $m(x_1)$, so the generation is solely conditioned on the text $z_p$ and the fine-grained conditions $z_f$.
To explore different aspects of our approach, we also conduct experiments using the prompted generation setup in selected cases. 
In this variant, we use the initial 3 seconds of the ground-truth recording as the speech prompt $m(x_1)$ to provide global information such as speaker, style, and environment noises, as opposed to the fine-grained information modeled by the proposed method.
The prompted generation setup enables us to study the disentanglement of global and fine-grained conditions in Voicebox Adapter. 
Implementation details for both setups follow Voicebox~\cite{le2023voicebox}.
\vspace{-2.5mm}
\subsection{Objective evaluation methods}
\vspace{-1mm}
\label{ssec:objective}
\subsubsection{Fine-grained controllability}
\vspace{-1mm}
We use automatic annotation tools to evaluate the adherence of the generated utterances to the fine-grained conditions.
For emphasis and laughter tasks, we follow the approach of the word emphasis detector in the EmphAssess benchmark~\cite{seyssel2023emphassess} to train the annotation models with our fine-grained conditional fine-tuning data.
For punctuation, we use the English-only Whisper-small model~\cite{radford2023robust} to add punctuation marks to the text transcription of generated speech by constraining the output of the Whisper decoder to either the next ground-truth text token or a punctuation mark in each decoding step.
The pseudo labels predicted by the annotators are compared with the original fine-grained conditions, and the results are reported in $F_1$ scores\footnotemark{} as in prior work~\cite{seyssel2023emphassess}.
As a sanity check, our annotation models achieve $F_1$ scores of 86.7\%, 92.4\%, and 62.1\% on the validation set of the emphasis, laughter, and punctuation tasks, respectively.



\vspace{-2mm}
\subsubsection{Intelligibility and speaker similarity}
\vspace{-1mm}
To assess whether fine-grained conditional fine-tuning affects the speech generation capabilities of our models, we use word error rate (WER) to evaluate the content correctness and intelligibility of utterances generated in the zero-shot scenario, and report speaker similarity (SIM-o) between the speech prompts and the utterances generated in the prompted scenario.
The setup of these evaluation methods follows Voicebox~\cite{le2023voicebox}.
\vspace{-2mm}
\subsection{Subjective evaluation methods}
\vspace{-1mm}
For the subjective evaluation, we recruit 34 trained subjects with audio relevant experience to rate the generated speech samples and report
our results using the 5-scale Mean Opinion Score (MOS).
In the fine-grained controllability MOS test (FC-MOS), listeners are instructed to focus on the alignment between the generated utterances and the fine-grained conditions, while disregarding speech quality, style, speaker characteristics, and other irrelevant aspects.
Conversely, in the quality MOS test (Q-MOS), listeners focus solely on the quality of the generated speech and disregard other factors.
In each subjective evaluation, we randomly select 200 samples from the evaluation set (with the exception of 100 samples for the emphasis task due to limited data availability) and collect 5 ratings for each sample. 
We use the recommended recipes from the CrowdMOS~\cite{ribeiro2021crowdmos} package to filter outliers and address inaccurate ratings, and report the averaged ratings with a 95\% confidence interval.

\begin{table}[b]
    \setlength{\tabcolsep}{1.25pt}
    \centering
    \vspace{-5mm}
    \caption{Performance of Voicebox Adapter with different adaptive modules on the fine-grained conditional generation tasks.}
    \vspace{-3mm}
    \footnotesize
    \begin{tabular}{l r r r r}
        \toprule
        \multirow{2}{*}{Adaptive Modules} & Punctuation & Emphasis & Laughter & \multirow{2}{*}{params.$^*$}\\
        \cmidrule(lr){2-4}
        & $F_1$(\%) & $F_1$(\%) & $F_1$(\%) & \\
        \cmidrule(lr){1-5}
        Sequential adapter & 63.6 & 66.9 & 39.9 & 2.4M\\
        Parallel adapter & 63.4 & 70.5 & 44.4 & 2.4M\\
        LoRA (self-attention only) & 63.5 & 74.7 & 44.5 & 3.5M\\
        \quad + bias-tuning & 63.8 & 75.6 & 50.4 & 3.6M\\
        LoRA (all linear layers) & 63.2 & 72.9 & 47.5 & 4.4M\\
        \bottomrule
    \end{tabular}
    \justify
    \vspace{-2.5mm}
    \scriptsize $^*$The number of parameters in the adaptive modules of the acoustic model. The pre-trained acoustic model has 93M parameters in total.
    \label{tab:adaptive_modules}
\end{table}

\begin{table*}[t]
    \scriptsize
    \setlength{\tabcolsep}{2pt}
    \centering
    \caption{Main results for the evaluations of fine-grained controllability, quality, intelligibility, and speaker similarity.}
    \vspace{-3mm}
    \begin{tabular}{l r r r r r r r r r r r r r r}
        \toprule
        \multirow{2}{*}{Models} & \multicolumn{3}{c}{Configuration} & \multicolumn{5}{c}{Punctuation (LibriTTS)} & \multicolumn{3}{c}{Emphasis (Expresso)} & \multicolumn{3}{c}{Laughter (Switchboard)}\\
        \cmidrule(lr){2-4}
        \cmidrule(lr){5-9}
        \cmidrule(lr){10-12}
        \cmidrule(lr){13-15}
        & PT & X-attn. & Adapt. & $F_1$(\%) & FC-MOS & WER(\%) & SIM-o & Q-MOS & $F_1$(\%) & FC-MOS & Q-MOS & $F_1$(\%) & FC-MOS & Q-MOS\\
        \cmidrule(lr){1-15}
        (a) Ground-truth & & & & 62.1 & 4.04$\pm$0.09 & 5.3 & 0.718 & 3.84$\pm$0.09 & 86.7 & 4.00$\pm$0.13 & 3.98$\pm$0.13 & 92.4 & 4.27$\pm$0.10 & 3.78$\pm$0.10\\
        \cmidrule(lr){1-15}
        (b) Voicebox Adapter & \check & \check & LoRA+BT & 63.8 & 4.18$\pm$0.07 & 3.2 & 0.593 & 3.88$\pm$0.09 & 75.6 & 3.57$\pm$0.18 & 3.82$\pm$0.13 & 50.4 & 2.15$\pm$0.17 & 3.67$\pm$0.09\\
        (c) Fine-tune all$^*$ & \check & \check & \cross & 63.5 & 4.20$\pm$0.07 & 3.3 & 0.568 & 3.89$\pm$0.08 & 71.9 & 3.62$\pm$0.16 & 3.82$\pm$0.13 & 57.5 & 2.55$\pm$0.19 & 3.64$\pm$0.09\\
        (d) No pre-training$^{*\dagger}$ & \cross & \check & \cross & 63.5 & 4.20$\pm$0.07 & 3.6 & 0.512 & 3.90$\pm$0.09 & 68.4 & 3.30$\pm$0.16 & 3.61$\pm$0.15 & 57.1 & 2.61$\pm$0.19 & 3.45$\pm$0.10\\
        \cmidrule(lr){1-15}
        (e) Voicebox$^{*\dagger}$ & \cross & \cross & \cross & 57.7 & 3.85$\pm$0.08 & 3.5 & 0.565 & 3.82$\pm$0.09 & 39.3 & 2.99$\pm$0.13 & 3.78$\pm$0.14 & 32.7 & 1.46$\pm$0.12 & 3.64$\pm$0.09\\
        \bottomrule
        \multicolumn{15}{l}{\scriptsize PT: pre-training; X-attn.: cross-attention; Adapt.: adaptive modules; BT: bias-tuning.}\\
        \multicolumn{15}{l}{\scriptsize $^*$For models without adaptive modules, we unfreeze all model parameters (except for T5) when training the models on the fine-grained conditioning datasets.}\\
        \multicolumn{15}{l}{\scriptsize $^\dagger$For models without pre-training, we train the acoustic and duration models for 150k/400k updates on the fine-grained conditioning datasets to ensure full convergence.}\\
    \end{tabular}
    \label{tab:main_f}
    \vspace{-5mm}
\end{table*}

\vspace{-2mm}
\section{Results}
\vspace{-1mm}
\subsection{Effectiveness of different efficient fine-tuning methods}
\vspace{-1mm}

We use three datasets to evaluate Voicebox Adapter on the fine-grained conditional generation tasks.
For the punctuation task, we use the \textit{test} set of the LibriTTS dataset~\cite{zen2019a}.
For the emphasis task, we randomly select 100 utterances with emphasis annotations from the Expresso dataset~\cite{communication2023seamless}. 
For the laughter task, we sample 27 speakers from the Switchboard dataset~\cite{godfrey1992switchboard} and use their utterances with laughter annotations for evaluation.

First, we would like to identify the optimal strategy for fine-tuning the pre-trained Voicebox model with fine-grained conditions.
Table~\ref{tab:adaptive_modules} shows the results of the efficient fine-tuning strategies we explore.
We find that parallel adapters outperform sequential adapters, which is consistent with prior research in both text and speech models~\cite{he2022towards, li2023modular}.
Notably, the LoRA + bias-tuning configuration demonstrates superior results across all tasks.
Considering that the differences in the parameter counts of the adaptive modules for each fine-tuning method are marginal --- accounting for less than 5\% of the pre-trained parameters --- we adopt LoRA + bias-tuning as our best setup and use it as our default configuration for the remainder of the paper.

\vspace{-3mm}
\subsection{Main results}
\vspace{-1mm}
To verify the effectiveness of Voicebox Adapter, we conduct a comprehensive comparison against the baseline Voicebox model and other alternative training configurations.
The results of our objective and subjective evaluations are shown in Table~\ref{tab:main_f}.

In the fine-grained controllability metrics $F_1$ and FC-MOS, we observe that models incorporating fine-grained conditioning inputs (rows (b), (c), and (d)) consistently outperform the baseline Voicebox model (row (e)).
Despite fine-tuning only a small portion of parameters, Voicebox Adapter (row (b)) shows performance comparable to fine-tuning the entire model (row (c)). 
The comparison between rows (b) and (d) highlights the importance of pre-training as Voicebox Adapter achieves comparable performance with significantly fewer fine-tuning iterations.

For models without applying efficient fine-tuning (row (c) and (d)), we find that pre-training does not make large differences to the punctuation and laughter tasks, but results in an improvement in the emphasis task, where the amount of data is limited.
This further shows the potential of the pre-training \& fine-tuning strategy in resource-constrained scenarios.

We notice that all models have worse performance on the Laughter task. 
This challenge may be attributed to limited presence of laughter in the pre-training data.
Our analysis shows that only 1.4\% of frames in the pre-training data trigger the laughter detector developed in section~\ref{ssec:objective}, whereas 6.3\% of the frames trigger the emphasis detector.
This discrepancy highlights the importance of adequate representation of various speech characteristics during pre-training.

The Q-MOS evaluation further shows that the efficient fine-tuning of Voicebox Adapter does not compromise the quality of the generated speech (row (b) vs. others). 
Further, the comparison between rows (b) (c) against row (d) shows that pre-training enhances the quality of generated speech, particularly on the Expresso and Switchboard datasets. 
The WER and SIM-o evaluations on the LibriTTS dataset validate that Voicebox Adapter retains the intelligibility and one-shot speech generation ability of Voicebox~\cite{le2023voicebox} and even exhibits improved performance, which may be attributed to the pre-training on large corpora.

\vspace{-2mm}
\subsection{Hyper-parameter and data ablation studies}
\begin{figure}[t]
    \centering
    \includegraphics[width=0.9\linewidth]{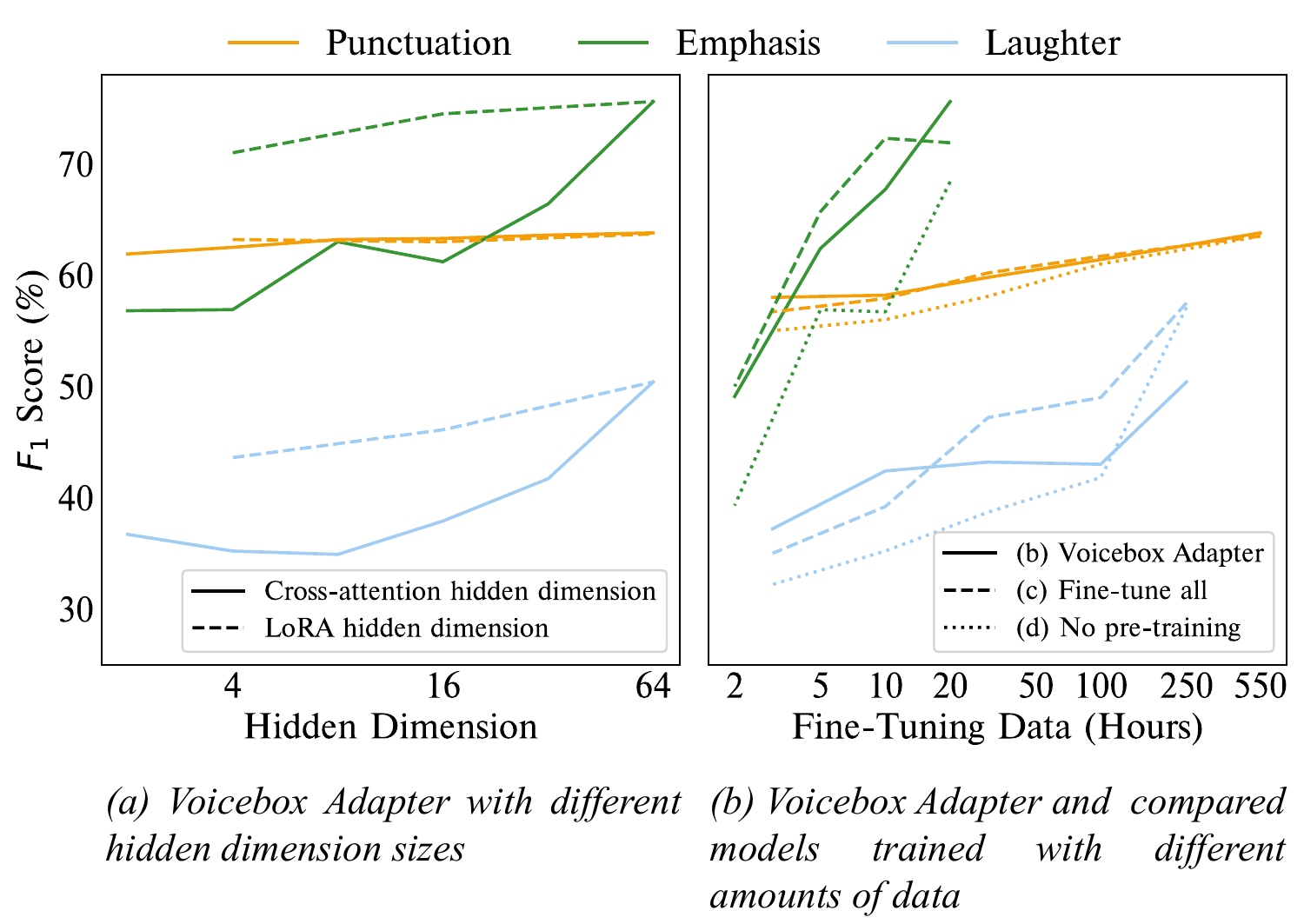}
    \vspace{-3mm}
    \caption{Performance of Voicebox Adapter and compared models with different hidden dimensions and data configurations.}
    \label{fig:ablation}
    \vspace{-6mm}
\end{figure}

\vspace{-1mm}
In Fig.~\ref{fig:ablation}(a), we present the results of models with different cross-attention dimensions and LoRA hidden dimensions $r$.
In general, we observe a positive correlation between the hidden dimensions and the performance of the models.
These findings contrast with the results of the original LoRA experiments on Large Language Models (LLMs)~\cite{hu2022lora}, which may be attributed to the size difference between Voicebox Adapter and state-of-the-art LLMs.
As reported, small intrinsic dimensions in large models~\cite{aghajanyan2021intrinsic} potentially lead to the good performance of LLMs with small LoRA hidden dimensions.
Given that Voicebox Adapter has less than 0.1\% of the parameters of GPT-3~\cite{brown2020language}, it is conceivable that a smaller $r$ could be more effective as the model size scales up.
However, this hypothesis requires further investigation, which we plan to explore more in future work.

In Figure~\ref{fig:ablation}(b), we fine-tune Voicebox Adapter on subsets of the fine-tuning data and compare their performance.
We find that the efficient fine-tuning method consistently achieves performance comparable to fine-tuning the entire model, regardless of the amount of data used in the fine-tuning process.
This observation holds true across different tasks, affirming the robustness of Voicebox Adapter across various data setups.



\vspace{-2mm}
\section{Conclusion}
\vspace{-1mm}
In this paper, we introduce Voicebox Adapter, which augments a pre-trained Voicebox speech generation model with fine-grained conditions. 
We inject fine-grained conditions into the Voicebox model through cross-attention modules and investigate various efficient fine-tuning methods to facilitate the integration of pre-trained parameters with newly introduced modules.
On the three fine-grained conditional generation tasks --- punctuation, emphasis, and laughter --- Voicebox Adapter successfully adheres to the fine-grained conditions while maintaining the naturalness and intelligibility of the generated speech.
In our subjective and objective evaluations, Voicebox Adapter shows superior performance over the baseline Voicebox model and achieves comparable performance to fine-tuning the entire model.
Future work involves extending the proposed method to more diverse fine-grained conditions, as well as addressing challenges in specific tasks such as laughter generation.


\clearpage

\bibliographystyle{IEEEtran}
\bibliography{refs}

\begin{thebibliography}{10}
\providecommand{\url}[1]{#1}
\csname url@samestyle\endcsname
\providecommand{\newblock}{\relax}
\providecommand{\bibinfo}[2]{#2}
\providecommand{\BIBentrySTDinterwordspacing}{\spaceskip=0pt\relax}
\providecommand{\BIBentryALTinterwordstretchfactor}{4}
\providecommand{\BIBentryALTinterwordspacing}{\spaceskip=\fontdimen2\font plus
\BIBentryALTinterwordstretchfactor\fontdimen3\font minus \fontdimen4\font\relax}
\providecommand{\BIBforeignlanguage}[2]{{%
\expandafter\ifx\csname l@#1\endcsname\relax
\typeout{** WARNING: IEEEtran.bst: No hyphenation pattern has been}%
\typeout{** loaded for the language `#1'. Using the pattern for}%
\typeout{** the default language instead.}%
\else
\language=\csname l@#1\endcsname
\fi
#2}}
\providecommand{\BIBdecl}{\relax}
\BIBdecl

\bibitem{yang2021superb}
S.~Yang, P.-H. Chi, Y.-S. Chuang, C.-I.~J. Lai, K.~Lakhotia, Y.~Y. Lin, A.~T. Liu, J.~Shi, X.~Chang, G.-T. Lin, T.-H. Huang, W.-C. Tseng, K.~Lee, D.-R. Liu, Z.~Huang, S.~Dong, S.-W. Li, S.~Watanabe, A.~Mohamed, and H.~Lee, ``{SUPERB}: Speech processing universal performance benchmark,'' in \emph{Interspeech}, 2021.

\bibitem{mohamed2022self}
A.~Mohamed, H.~Lee, L.~Borgholt, J.~D. Havtorn, J.~Edin, C.~Igel, K.~Kirchhoff, S.-W. Li, K.~Livescu, L.~Maaløe, T.~N. Sainath, and S.~Watanabe, ``Self-supervised speech representation learning: A review,'' \emph{IEEE Journal of Selected Topics in Signal Processing}, vol.~16, no.~6, pp. 1179--1210, 2022.

\bibitem{baevski2020wav2vec}
A.~Baevski, Y.~Zhou, A.~Mohamed, and M.~Auli, ``wav2vec 2.0: A framework for self-supervised learning of speech representations,'' in \emph{NeurIPS}, 2020.

\bibitem{hsu2021hubert}
W.-N. Hsu, B.~Bolte, Y.-H.~H. Tsai, K.~Lakhotia, R.~Salakhutdinov, and A.~Mohamed, ``{H}u{BERT}: Self-supervised speech representation learning by masked prediction of hidden units,'' \emph{IEEE/ACM Transactions on Audio, Speech, and Language Processing}, vol.~29, pp. 3451--3460, 2021.

\bibitem{liu2024generative}
A.~H. Liu, M.~Le, A.~Vyas, B.~Shi, A.~Tjandra, and W.-N. Hsu, ``Generative pre-training for speech with flow matching,'' in \emph{ICLR}, 2024.

\bibitem{lakhotia2021on}
K.~Lakhotia, E.~Kharitonov, W.-N. Hsu, Y.~Adi, A.~Polyak, B.~Bolte, T.-A. Nguyen, J.~Copet, A.~Baevski, A.~Mohamed, and E.~Dupoux, ``On generative spoken language modeling from raw audio,'' \emph{Transactions of the Association for Computational Linguistics}, vol.~9, pp. 1336--1354, 2021.

\bibitem{borsos2023audio}
Z.~Borsos, R.~Marinier, D.~Vincent, E.~Kharitonov, O.~Pietquin, M.~Sharifi, D.~Roblek, O.~Teboul, D.~Grangier, M.~Tagliasacchi, and N.~Zeghidour, ``{AudioLM}: A language modeling approach to audio generation,'' \emph{IEEE/ACM Transactions on Audio, Speech, and Language Processing}, vol.~31, pp. 2523--2533, 2023.

\bibitem{nguyen2023generative}
T.~A. Nguyen, E.~Kharitonov, J.~Copet, Y.~Adi, W.-N. Hsu, A.~Elkahky, P.~Tomasello, R.~Algayres, B.~Sagot, A.~Mohamed, and E.~Dupoux, ``Generative spoken dialogue language modeling,'' \emph{Transactions of the Association for Computational Linguistics}, vol.~11, pp. 250--266, 2023.

\bibitem{wang2023neural}
C.~Wang, S.~Chen, Y.~Wu, Z.~Zhang, L.~Zhou, S.~Liu, Z.~Chen, Y.~Liu, H.~Wang, J.~Li, L.~He, S.~Zhao, and F.~Wei, ``Neural codec language models are zero-shot text to speech synthesizers,'' \emph{preprint arXiv:2301.02111}, 2023.

\bibitem{le2023voicebox}
M.~Le, A.~Vyas, B.~Shi, B.~Karrer, L.~Sari, R.~Moritz, M.~Williamson, V.~Manohar, Y.~Adi, J.~Mahadeokar, and W.-N. Hsu, ``Voicebox: Text-guided multilingual universal speech generation at scale,'' in \emph{NeurIPS}, 2023.

\bibitem{chen2019sample}
Y.~Chen, Y.~M. Assael, B.~Shillingford, D.~Budden, S.~E. Reed, H.~Zen, Q.~Wang, L.~C. Cobo, A.~Trask, B.~Laurie, {\c{C}}.~G{\"{u}}l{\c{c}}ehre, A.~van~den Oord, O.~Vinyals, and N.~de~Freitas, ``Sample efficient adaptive text-to-speech,'' in \emph{ICLR}, 2019.

\bibitem{chen2021adaspeech}
M.~Chen, X.~Tan, B.~Li, Y.~Liu, T.~Qin, S.~Zhao, and T.~Liu, ``{AdaSpeech}: Adaptive text to speech for custom voice,'' in \emph{ICLR}, 2021.

\bibitem{yan2021adaspeech}
Y.~Yan, X.~Tan, B.~Li, G.~Zhang, T.~Qin, S.~Zhao, Y.~Shen, W.-Q. Zhang, and T.-Y. Liu, ``Adaptive text to speech for spontaneous style,'' in \emph{Interspeech}, 2021.

\bibitem{huang2022generspeech}
R.~Huang, Y.~Ren, J.~Liu, C.~Cui, and Z.~Zhao, ``Generspeech: Towards style transfer for generalizable out-of-domain text-to-speech,'' in \emph{NeurIPS}, 2022.

\bibitem{houlsby2019parameter}
N.~Houlsby, A.~Giurgiu, S.~Jastrzebski, B.~Morrone, Q.~De~Laroussilhe, A.~Gesmundo, M.~Attariyan, and S.~Gelly, ``Parameter-efficient transfer learning for {NLP},'' in \emph{ICML}, 2019.

\bibitem{hu2022lora}
E.~J. Hu, Y.~Shen, P.~Wallis, Z.~Allen{-}Zhu, Y.~Li, S.~Wang, L.~Wang, and W.~Chen, ``{LoRA}: Low-rank adaptation of large language models,'' in \emph{ICLR}, 2022.

\bibitem{gao2023llama}
P.~Gao, J.~Han, R.~Zhang, Z.~Lin, S.~Geng, A.~Zhou, W.~Zhang, P.~Lu, C.~He, X.~Yue, H.~Li, and Y.~Qiao, ``{LLaMA-Adapter V2}: Parameter-efficient visual instruction model,'' \emph{preprint arXiv:2304.15010}, 2023.

\bibitem{lipman2023flow}
Y.~Lipman, R.~T.~Q. Chen, H.~Ben-Hamu, M.~Nickel, and M.~Le, ``Flow matching for generative modeling,'' in \emph{ICLR}, 2023.

\bibitem{raffel2020exploring}
C.~Raffel, N.~Shazeer, A.~Roberts, K.~Lee, S.~Narang, M.~Matena, Y.~Zhou, W.~Li, and P.~J. Liu, ``Exploring the limits of transfer learning with a unified text-to-text transformer,'' \emph{Journal of Machine Learning Research}, vol.~21, no. 140, pp. 1--67, 2020.

\bibitem{he2022towards}
J.~He, C.~Zhou, X.~Ma, T.~Berg{-}Kirkpatrick, and G.~Neubig, ``Towards a unified view of parameter-efficient transfer learning,'' in \emph{ICLR}, 2022.

\bibitem{dettmers2023q}
T.~Dettmers, A.~Pagnoni, A.~Holtzman, and L.~Zettlemoyer, ``{QLoRA}: Efficient finetuning of quantized {LLMs},'' in \emph{NeurIPS}, 2023.

\bibitem{kong2020hifi}
J.~Kong, J.~Kim, and J.~Bae, ``{HiFi-GAN}: Generative adversarial networks for efficient and high fidelity speech synthesis,'' in \emph{NeurIPS}, 2020.

\bibitem{seyssel2023emphassess}
M.~de~Seyssel, A.~D'Avirro, A.~Williams, and E.~Dupoux, ``{EmphAssess}: {A} prosodic benchmark on assessing emphasis transfer in speech-to-speech models,'' \emph{preprint arXiv:2312.14069}, 2023.

\bibitem{radford2023robust}
A.~Radford, J.~W. Kim, T.~Xu, G.~Brockman, C.~Mcleavey, and I.~Sutskever, ``Robust speech recognition via large-scale weak supervision,'' in \emph{ICML}, 2023.

\bibitem{ribeiro2021crowdmos}
F.~Ribeiro, D.~Florêncio, C.~Zhang, and M.~Seltzer, ``{CROWDMOS}: An approach for crowdsourcing mean opinion score studies,'' in \emph{ICASSP}, 2011.

\bibitem{zen2019a}
H.~Zen, V.~Dang, R.~Clark, Y.~Zhang, R.~J. Weiss, Y.~Jia, Z.~Chen, and Y.~Wu, ``{LibriTTS}: A corpus derived from {LibriSpeech} for text-to-speech,'' in \emph{Interspeech}, 2019.

\bibitem{communication2023seamless}
S.~Communication, L.~Barrault, Y.-A. Chung, M.~C. Meglioli, D.~Dale, N.~Dong, M.~Duppenthaler, P.-A. Duquenne, B.~Ellis, H.~Elsahar, J.~Haaheim, J.~Hoffman, M.-J. Hwang, H.~Inaguma, C.~Klaiber, I.~Kulikov, P.~Li, D.~Licht, J.~Maillard, R.~Mavlyutov, A.~Rakotoarison, K.~R. Sadagopan, A.~Ramakrishnan, T.~Tran, G.~Wenzek, Y.~Yang, E.~Ye, I.~Evtimov, P.~Fernandez, C.~Gao, P.~Hansanti, E.~Kalbassi, A.~Kallet, A.~Kozhevnikov, G.~M. Gonzalez, R.~S. Roman, C.~Touret, C.~Wong, C.~Wood, B.~Yu, P.~Andrews, C.~Balioglu, P.-J. Chen, M.~R. Costa-jussà, M.~Elbayad, H.~Gong, F.~Guzmán, K.~Heffernan, S.~Jain, J.~Kao, A.~Lee, X.~Ma, A.~Mourachko, B.~Peloquin, J.~Pino, S.~Popuri, C.~Ropers, S.~Saleem, H.~Schwenk, A.~Sun, P.~Tomasello, C.~Wang, J.~Wang, S.~Wang, and M.~Williamson, ``Seamless: Multilingual expressive and streaming speech translation,'' \emph{preprint arXiv:2312.05187}, 2023.

\bibitem{godfrey1992switchboard}
J.~Godfrey, E.~Holliman, and J.~McDaniel, ``{SWITCHBOARD}: Telephone speech corpus for research and development,'' in \emph{ICASSP}, 1992.

\bibitem{li2023modular}
Q.~Li, B.~Li, D.~Hwang, T.~Sainath, and P.~M. Mengibar, ``Modular domain adaptation for conformer-based streaming {ASR},'' in \emph{InterSpeech}, 2023.

\bibitem{aghajanyan2021intrinsic}
A.~Aghajanyan, S.~Gupta, and L.~Zettlemoyer, ``Intrinsic dimensionality explains the effectiveness of language model fine-tuning,'' in \emph{ACL-IJCNLP}, 2021.

\bibitem{brown2020language}
T.~Brown, B.~Mann, N.~Ryder, M.~Subbiah, J.~D. Kaplan, P.~Dhariwal, A.~Neelakantan, P.~Shyam, G.~Sastry, A.~Askell, S.~Agarwal, A.~Herbert-Voss, G.~Krueger, T.~Henighan, R.~Child, A.~Ramesh, D.~Ziegler, J.~Wu, C.~Winter, C.~Hesse, M.~Chen, E.~Sigler, M.~Litwin, S.~Gray, B.~Chess, J.~Clark, C.~Berner, S.~McCandlish, A.~Radford, I.~Sutskever, and D.~Amodei, ``Language models are few-shot learners,'' in \emph{NeurIPS}, 2020.

\end{thebibliography}

\end{document}